\documentclass[final,5p,times]{elsarticle}

\usepackage{amssymb,amsmath,nccmath}

\usepackage{graphicx,natbib}

\newcommand{\Hmax}{$H_{\rm max}$}
\newcommand{\DS}{$\Delta S$}
\newcommand{\Tad}{$\Delta T_{\rm ad}$}
\newcommand{\DSmax}{$\Delta S^{\rm max}$}
\newcommand{\Tadmax}{$\Delta T^{\rm max}_{\rm ad}$}

\newcommand{\beq}{\begin{equation}}
\newcommand{\eeq}{\end{equation}}
\newcommand{\bdis}{\begin{displaymath}}
\newcommand{\edis}{\end{displaymath}}
\newcommand{\bea}{\begin{eqnarray}}
\newcommand{\eea}{\end{eqnarray}}
\newcommand{\barr}{\begin{array}}
\newcommand{\earr}{\end{array}}
\newcommand{\bfig}{\begin{figure}[!]}
\newcommand{\efig}{\end{figure}}

\journal{Scripta Materialia}

\begin{document}

\begin{frontmatter}


 \title{Title}
 \author{Karl G. Sandeman}
 \ead{k.sandeman@imperial.ac.uk}
 \ead[url]{http://www.imperial.ac.uk/people/k.sandeman}
 \address{Department of Physics, Blackett Laboratory, Imperial College London, London SW7 2AZ, United Kingdom}

\title{Magnetocaloric materials: the search for new systems}

\begin{abstract}
The prospect of efficient solid state refrigeration at room temperature is driving research into magnetic cooling engine design and magnetic phase transition-based refrigerants. In this Viewpoint an Ashby-style map of magnetic refrigerant properties is constructed, comparing popular materials with limits derived from an idealised first order transition model. This comparison demonstrates the potential for new magnetocaloric material systems to be established through structural control and optimisation at the atomic-, nano- and micro-scale.
\end{abstract}


\end{frontmatter}


\section{Introduction}
\label{Intro}
It is often stated that room temperature magnetic cooling is motivated by two factors: firstly, the elimination of refrigerants that are greenhouse gases and secondly, the potential system efficiency gain.  The former is perhaps more obvious since solid refrigerants are used in magnetic cooling; the only fluids present are non-volatile heat exchange media.  The gain in final system efficiency should be carefully stated.  It will probably be most apparent at low cooling powers.  In that regime, the system efficiency of conventional refrigerators is compromised by the fact that manufacturers typically use a relatively efficient, high power compressor but operate it for only a fraction of the available time~\cite{sandeman_2011a}.  A variable speed compressor is an unattractively expensive route to efficiency for many low power applications.

There has therefore been a recent growth in research of room temperature magnetic cooling engines and magnetic refrigerants.  Regarding the engines, published studies have addressed regenerator geometries~\cite{kitanovskii_2009a,engelbrecht_2011a}, refrigerant morphologies~\cite{bjork_2011a,pryds_2011a}, cycle frequency~\cite{richard_2004a,kuzmin_2007a,russek_2010a} and magnetic field provision~\cite{bjork_2010a}.  Meanwhile, refrigerant research has examined both well-known and novel magnetic phase transition systems, considering principally the composition dependence of magnetothermal properties.  These are typically the isothermal entropy change $\Delta S$, or adiabatic temperature change, {\Tad}, induced by a magnetic field as a function of temperature around the phase transition.  Other, less frequently explored (but extremely relevant) properties are the origin and size of magnetic and thermal hysteresis, thermal conductivity, corrosion properties and machinability.  The results of cooling engine research are also beginning to influence refrigerant morphology and microstructure and their influence on magnetocaloric properties has recently been studied~\cite{sseec18month,kuhn_2011a}.  From this overview we understand that the building of a magnetic cooling engine is a highly interdisciplinary challenge.

This Viewpoint focuses on magnetic refrigerants. The properties of magnetocaloric materials are sensitive to changes in structure at all scales; from the atomic and nano-scale (exchange interactions) to the micro-scale (phase content, internal strain and homogeneity) and the macro-scale (shape, demagnetisation, pressing).  Such sensitivity is similar to the situation in permanent magnets where, for example, coercivity is limited by the intrinsic anisotropy field but is greatly affected (lowered) by a range of extrinsic properties at different length scales.  Structural sensitivity is heightened in room temperature magnetocaloric materials since cooling is associated with a phase transition.  

A ``new" refrigerant system can therefore arise from a change in structure at any scale, including a change in the microstructure of a fixed bulk composition.  Indeed, part of this Viewpoint will argue that near-term deployment of magnetocaloric materials will be  accelerated by intelligent synthesis and microstructuring of known material compositions.  This is one aspect of the search for new systems.  At the same time, research into the fundamental properties of magnetocaloric materials is yielding new compositions, new magnetocaloric effect mechanisms and a range of advanced characterisation tools that both advance our knowledge of this sensitive class of magnetic materials and provide invaluable feedback on refrigerant design.

I  therefore discuss refrigerant properties under three broad headings: fundamental magnetothermal properties, refrigerant comparison, and challenges for material deployment.   Each impacts the near-term feasibility of efficient magnetic cooling and the prospect of making the cost of the refrigerant  a trivial consideration. 
\section{Fundamental magnetothermal properties}
\label{MR_use}
Magnetic refrigerants need to have a large magnetocaloric effect (MCE).  If we use continuous thermodynamics, then the isothermal application of a magnetic field, {\Hmax} causes a change of entropy, {\DS}:
\begin{equation}
\Delta S(T,H_{\rm max}) = \int_{0}^{H_{\rm max}} \biggl({\partial M \over \partial T}\biggr)_{H} dH \, .
\label{DSEq}
\end{equation}
If the field is applied adiabatically it causes a change of temperature of the material, {\Tad}:
\begin{equation}
\Delta T_{\rm ad}(T,H_{\rm max}) = \int_{0}^{H_{\rm max}} {T\over C_{p}(T,H)}\biggl({\partial M \over \partial T}\biggr)_{H} dH \,.
\label{DTEq}
\end{equation}
Note that in the former case the applied field changes the total entropy of the material.  If there are simultaneous, reversible changes in lattice entropy, for example at a coupled magneto-elastic transition, these are also captured by the above expression~\cite{mukherjee_2011a}.  Similarly, if the magnetisation vector is rotated by the applied field, causing a change in the free energy due to magnetocrystalline anisotropy, the resulting temperature or entropy change is also contained in Eqs.~\ref{DSEq} and~\ref{DTEq} above.  

We can thus use a single analysis framework to consider all magnetic transitions, such as: Curie transitions involving magnetic disordering of a ferromagnet, order-order transitions such as those seen in low anisotropy antiferromagnets, and spin reorientation transitions in which the direction(s) of easy magnetisation change as a function of temperature.  Before considering the prospects for new systems, I review and develop arguments concerning the limits of the properties in Eqs.~\ref{DSEq} and~\ref{DTEq}.

\subsection{The theoretical limit of {\Tad}}
\label{sec_Tad}
Prototype room temperature regenerator designs have traditionally used Gd as the refrigerant~\cite{yu_2010a}.  More recently,  La-Fe-Si~\cite{russek_2010a}, (La,Ca)MnO$_{3}$~\cite{pryds_2011a} have been trialled, and Mn-Fe-P-based refrigerants have been proposed~\cite{MnFeP-patent}.  All of these can have {\Tad} of at least 1~K when $\mu_{0}${\Hmax} is 1~Tesla (achievable with a permanent magnet).  This may be taken as a minimum requirement for {\Tad} when considering a magnetic refrigerant for application, due to the presence of several loss mechanisms in the regenerator.  They include: heat flow between the solid refrigerant and the liquid heat exchange medium, viscous flow entropy generation in the liquid heat exchanger and thermal backflow in the refrigerant bed.  Their presence influences our judgement of the usefulness of materials, and the form that they might take in the regenerator.  In addition, the thermodynamics of the magnetic cooling system as a whole is important, and has most recently been expressed in terms of its exergetic efficiency.  This is a measure of the energy available for cooling, and depends on the temperature span of the device~\cite{rowe_2009a}.  

The field variation of {\DS}$(T)$ and {\Tad}$(T)$ have been the subject of a number of recently studies, investigating either their peak values at each field, $\Delta S^{\rm max}$~\cite{lyubina_2011a} or $\Delta T_{\rm ad}^{\rm max}$~\cite{kuzmin_2011a}, or using scaling laws to obtain universality curves for the behaviour of {\DS}$(T,H)$~\cite{franco_2010a,dong_2007a,amaral_2007a}.   Most of these studies have been on materials with continuous phase transitions; the scaling behaviour of first order materials is less predictable. However, it is worth noting that the peak {\DS}$(T)$ data for some low-hysteresis, first order La-Fe-Si materials have been successfully fitted using the same fourth order Landau theory as for continuous phase transition compounds, with a inhomogeneous distribution of Curie temperatures~\cite{lyubina_2011a}.  Further work on this regime is required, as it is interesting both for fundamental understanding and for the promise of low hysteresis first order materials with high {\DS}.

\begin{figure}
\includegraphics[width=\columnwidth]{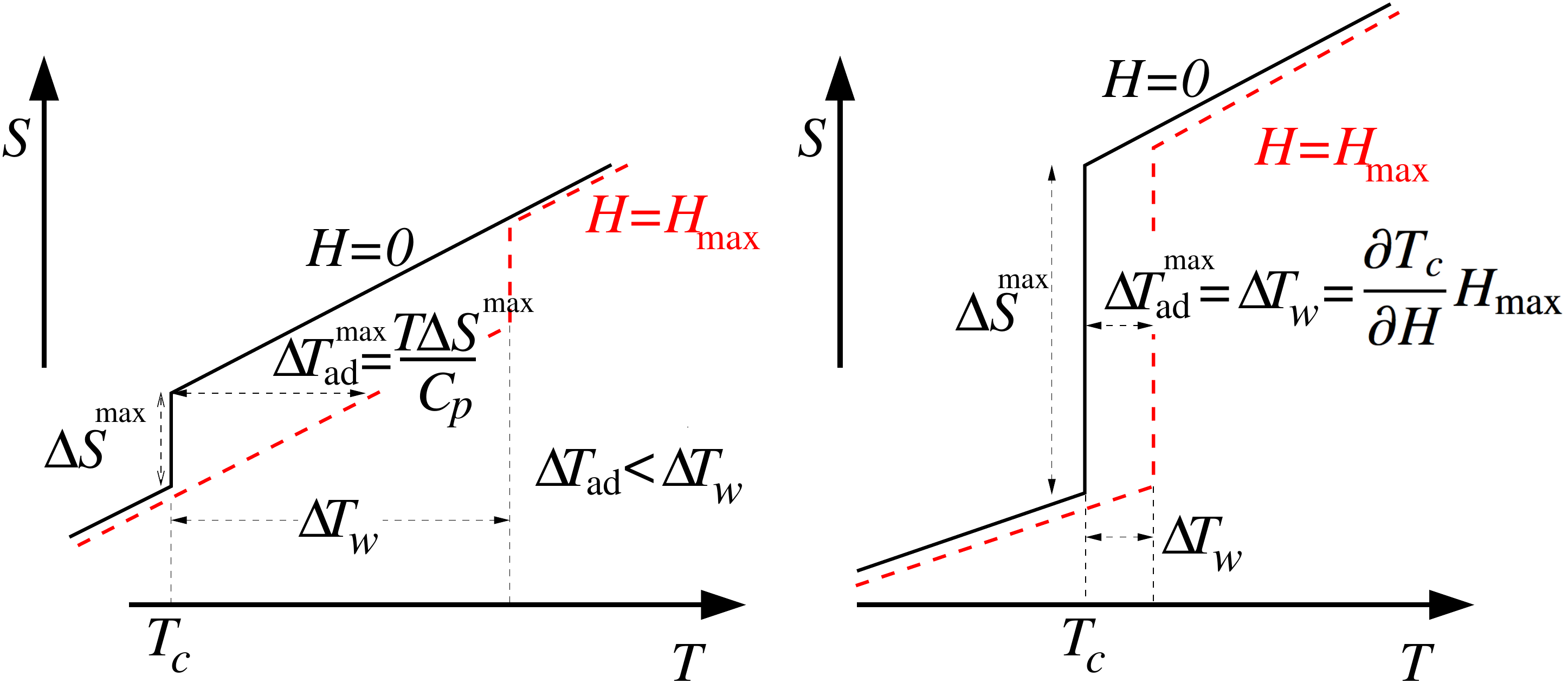}
\caption{\label{DTmaxschematicFig} If the rate of change of a magnetic phase transition temperature with magnetic field, $\partial T_{c} / \partial H$, is large (left) or small (right) then two quite different entropy vs. temperature scenarios arise.  In between these cases there is an optimal value of $\partial T_{c} / \partial H=\eta$ that optimises {\Tadmax}, leading to Eq.~\ref{eq_DTmax}.  In each case the black line is in zero field; the red dashed line in a finite field {\Hmax}.}
\end{figure}

Since the above studies concentrate predominantly on continuous phase transitions, gadolinium can appear to be the ``most efficient" refrigerant~\cite{kuzmin_2011a}.  A similar conclusion was reached by Zverev et al., in considering the maximum possible {\Tad} near a Curie transition~\cite{zverev_2010a}.  The authors started from the approximation that, within a small temperature interval, $({\partial M \over \partial T})_{H}$ is constant, and showed that the optimum value of {\Tadmax} is 
\begin{equation}
(\Delta T_{\rm ad}^{\rm max})|_{\rm optimal}=\bigl({M_{\rm sat}T H_{\rm max} \over C_{p}} \bigr)^{1\over 2} \, .
\label{eq_DTmax}
\end{equation}
This relation can also be derived by varying $\partial T_{c} / \partial H$, the rate at which a general transition temperature $T_{c}$ changes with field.  The importance of this parameter was first highlighted by Tishin~\cite{tishin_2003a} who showed that the experimentally observed relative cooling power (RCP) scales linearly with $\partial T_{c} / \partial H$, where RCP was defined according to the method of Gschneidner and Pecharsky as {\DS}$\Delta T_{1/2}$, $\Delta T_{1/2}$ being the full-width-half-maximum (FWHM) of the $\Delta S(T)$ response curve~\cite{gschneidner_2000a}.  If we consider an idealised first order transition, the aforementioned approximation applies in that the differential $({\partial M \over \partial T})_{H}$ is assumed infinite.  Then Eq.~\ref{eq_DTmax} may be obtained by invoking the magnetothermal ``sum rule"~\cite{bennett_1992a}.  This states that the isothermal entropy change is bounded by the total change in magnetisation and the applied field:
\begin{equation}
\int_{0}^{\infty} \Delta S(T,H_{\rm max}) dT=M_{\rm sat} H_{\rm max}\, ,
\label{SumRuleEq}
\end{equation}
where $M_{\rm sat}$ is the saturation magnetisation.  Since {\Tad} is the temperature difference between two isofield $S(T)$ curves, the bound on entropy change yields a bound on {\Tad}.  Two limiting scenarios, described in~\cite{zverev_2010a} arise from extremes in $\partial T_{c} / \partial H$ as shown schematically in Figure~\ref{DTmaxschematicFig}.   If $\partial T_{c} / \partial H$  is small, {\Tad} is equal to the temperature width, $\Delta T_w$, of {\DS}, which is $\partial T_{c} / \partial H \times H_{\rm max}$ and is therefore linear in $\partial T_{c} / \partial H$.  This is the regime commented on in Figure~5 of Ref.~\cite{pecharsky_2006a}.   However if  $\partial T_{c} / \partial H$  is large, the adiabatic temperature change is determined by the heat capacity, $C_p$, of the material away from the phase transition and the isothermal entropy change, such that {\Tad}$=T(${\DS}$/C_{p})$. In that case {\Tad} varies as ($\partial T_{c} / \partial H)^{-1}$.  From Eq.~\ref{SumRuleEq} and Figure~\ref{DTmaxschematicFig}:
\begin{eqnarray}
\Delta T_{\rm ad}^{\rm max}={T \Delta S^{\rm max} \over C_{p}} = {T \over C_{p}}{M_{\rm sat}H_{\rm max}\over (\partial T_{c} / \partial H)H_{\rm max}}\nonumber \\ 
= {T \over C_{p}}{M_{\rm sat}\over (\partial T_{c} / \partial H)} \, .
\end{eqnarray}
The maximum adiabatic temperature change occurs when the two opposing trends in {\Tadmax} vs. $\partial T_{c} / \partial H$ coincide:
\begin{equation}
{T \over C_{p}}{M_{\rm sat}\over (\partial T_{c} / \partial H)}={\partial T_{c} \over \partial H} H_{\rm max} \Rightarrow {\partial T_{c} \over \partial H}=\bigl({M_{\rm sat}T \over C_{p} H_{\rm max}} \bigr)^{1\over 2}=\eta \, ,
\label{eq_dTdHmax}
\end{equation}
which leads to Eq.~\ref{eq_DTmax}.  


\subsection{The cooling power limit}
We now examine the popular concept of Òrelative cooling powerÓ (RCP), which involves the available temperature range of {\DS}, rather than just its peak value, $\Delta S^{\rm max}$.  The RCP is calculated in different ways, usually depending on the temperature range over which the entropic response is integrated.  Wood and Potter, in their 1985 work on the magnetocaloric effect at low temperature~\cite{wood_1985a}, proposed an energy measure that is the maximal area of the product {\DS}$\times (T_{\rm hot}-T_{\rm cold})$ where $T_{\rm hot}$ and $T_{\rm cold}$ are the hot and cold working temperatures, respectively.   If we consider the regime $\partial T_{c} / \partial H<\eta$, then in the simple model presented here, the entropy change is uniform at all temperatures within $\Delta T_w$.  In that case, the relative cooling power of the material according to Wood and Potter is given by the sum rule in Eq.~(4) and simply equals $M_{\rm sat}${\Hmax}.   Theoretically, all materials with equal saturation magnetisation are equivalent by this measure.    In practice, $\Delta S(T)$ curves are not idealised and are more rounded.  Therefore an alternative measure, as described above, is {\DS}$\Delta T_{1/2}$ where $\Delta T_{1/2}$ is the FWHM of the $\Delta S(T)$ response curve~\cite{gschneidner_2000a}.  This gives some measure of the ``useful" entropy within the $\Delta S(T)$ curve, as long as the $T_{1/2}$ points relate to temperatures at which efficient heat exchange is possible; in other words that {\Tad} is sufficiently high.  If this is not the case and the $T_{1/2}$ points are defined too generously, refrigerants may appear similar merely because of their comparable $M_{\rm sat}$ values.

The {\it maximum} RCP is clearly $M_{\rm sat} H_{\rm max}$ irrespective of definition, if $\Delta T_w$ is sufficiently narrow (i.e. if $\partial T_{c} / \partial H < \eta$ given by Eq.~(6)).  However, let us use the results above to calculate $\mid${\DSmax}$\mid$$\mid${\Tadmax}$\mid$ since this is relevant to the Ashby map in Fig.~2 below.  (Here magnitudes are taken to account for inverse MCE materials and $T_c$ represents a generalised transition temperature.)  From the idealised first order material model, there are two scenarios:
\begin{eqnarray}
\mid\Delta T_{\rm ad}^{\rm max}\Delta S^{\rm max}\mid= \mid \Delta S^{\rm max}\Delta T_{w}\mid=M_{\rm sat} H_{\rm max} \,\,{\rm for} \,\,{\partial T_{c} \over \partial H}<\eta \,; \nonumber \\
\mid\Delta T_{\rm ad}^{\rm max}\Delta S^{\rm max}\mid = {T\Delta S^{2}\over C_{p}}= {T M_{\rm sat}\over C_{p} (\partial T_{c} / \partial H)^{2} } \,\,{\rm for} \,\, {\partial T_{c} \over \partial H}>\eta  \,.
\label{eq_TadHighdTdH}
\end{eqnarray}
High RCP values calculated by either of the usual entropy-based methods can obscure the impracticality of magnetothermal properties at $T_{\rm hot}$ and $T_{\rm cold}$, since a minimum {\Tad} is required for regenerator efficiency.  However $\mid${\DSmax}$\mid$$\mid${\Tadmax}$\mid$ is a quantity that is maximised by simultaneously large values of {\Tad} and {\DS}.
\section{Magnetic refrigerant comparison}
\label{ComparisonSection}
Much of the magnetic refrigerant literature has focussed on the field-induced isothermal entropy change, {\DS}.  This is probably because of the relative ease with which magnetisation measurements can be made, from which {\DS} can be obtained using Eq.~\ref{DSEq} under the correct circumstances.    As a result, recent reviews~\cite{tishin_2007a,yu_2003a,bruck_2005a,liumin_2009a,shen_2009a,phan_2007a} focus on {\DS}.   A plot of {\DS} vs. Curie temperature for $\mu_{0}${\Hmax}=1 or 2~Tesla in a large range of material systems arguably reveals few clear features (Figure~5 of~\cite{phan_2007a}; Figure~4 of~\cite{bruck_2005a}).  Meanwhile there are few straightforward comparisons of {\Tad} for different materials in the literature, especially at permanent magnet field strengths.

The situation is changing with the development of several new facilities around the world for measuring {\Tad}, employing either a contacted or a contactless approach~\cite{dankov_1997a, barcza_2011b}. Also, even if {\DS} and {\Tad}$(T)$ data have historically been shown in 2~Tesla and 5~Tesla field changes, the advances in scaling theory mentioned in section~\ref{sec_Tad}, in particular for continuous phase transition materials, allow a rescaling of some data down to permanent magnet field strengths.   The question remains how to best use these data.  The identification of a figure of merit for magnetic refrigerants is not as clear as for thermoelectrics where $ZT=\sigma S^{2} T / \kappa$, is used ($\sigma$ is the electrical conductivity, $\kappa$ the thermal conductivity, and $S$ is the thermopower).  Even for a thermoelectric device, efficiency is not solely determined by $ZT$.  It is known for thermoelectrics that the operating temperature range is key, and research seeks to optimise this quantity, whilst maintaining thermodynamic compatibility of the materials that make up the device~\cite{snyder_2008a}. For magnetic refrigerants, the question of a suitable figure of merit has been raised at symposia over the past 5 years.   Ultimately it is impossible to separate the refrigerant from the device; a good material implemented poorly is no solution.  However, the ``best" material will allow the most headroom for device-centred losses.   One goal might therefore be to obtain a cost-trivial material with optimal properties, even over a small temperature range, so that multiple compositions can give excellent performance over the range required in application.

\begin{figure}
\includegraphics[width=\columnwidth]{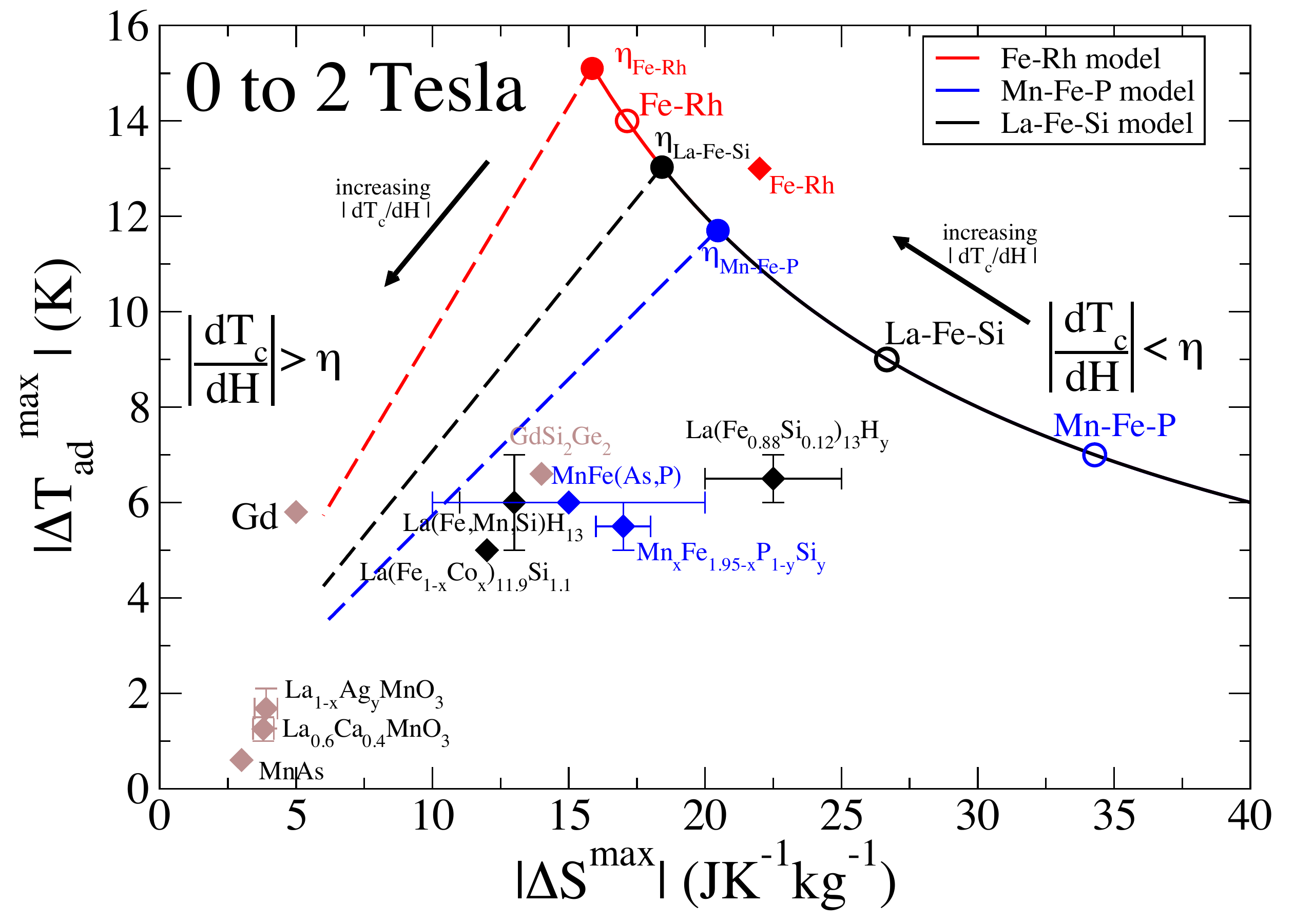}
\caption{\label{DTDSFig}$\mid${\Tadmax}$\mid$ vs. $\mid${\DSmax}$\mid$  for several room temperature refrigerants in a field change of 2~Tesla.  Data are taken from: Gd~\cite{gschneidner_1999a}; Gd$_5$Si$_2$Ge$_2$~\cite{pecharsky_2003a,gschneidner_2000b}; Fe-Rh~\cite{annaorazov_1992a,annaorazov_1996a}; La(Fe$_{1-x}$Co$_{x}$)$_{11.9}$Si$_{1.1}$~\cite{hu_2005a}; La(Fe$_{0.88}$Si$_{0.12}$)$_{13}$H$_{y}$~\cite{fujita_2003a}; LaFe$_{11.74-y}$Mn$_{y}$Si$_{1.26}$H$_{1.53}$~\cite{morrison_private}; La(Fe$_{1-x}$Mn$_{x}$)$_{11.7}$Si$_{1.3}$H$_{y}$~\cite{wang_2003a}; MnAs~\cite{tocado_2006a};  MnFe(P,As)~\cite{bruck_2005a}; Mn$_{x}$Fe$_{1.95-x}$P$_{1-y}$Si$_{y}$~\cite{dung_2011a}; La$_{0.6}$Ca$_{0.4}$MnO$_{3}$~\cite{dinesen_2002a};  La$_{1-x}$Ag$_{y}$MnO$_{3}$~\cite{kamilov_2007a}.  Diamonds are experimental data where error bars arise from the variation in magnetothermal properties across a series of compositions.  Theoretical {\Tadmax} and {\DSmax} are shown for $\partial T_{c} / \partial H$ values that are: $<\eta$ (solid line), $>\eta$ (dashed line) and at optimum $\partial T_{c} / \partial H=\eta$ (filled circles) for Fe-Rh, La-Fe-Si and Mn-Fe-P-based materials. Maximum theoretical values of {\Tadmax} and {\DSmax} using experimental values of $\partial T_{c} / \partial H$ ($<\eta$) are given by hollow circles.
}
\end{figure}

Here I therefore plot  $\mid${\Tadmax}$\mid$ vs. $\mid${\DSmax}$\mid$ in Figure~\ref{DTDSFig} for materials with MCEs in the room temperature range (270~K to 320~K) in order to construct an Ashby map~\cite{ashby_2005a} for magnetic refrigerants.  An ideal material will occupy the upper right area.  I have focused on popular materials with a reversible MCE with the exception of those that, as now discussed, inform future work despite their history-dependent MCE (MnAs~\cite{tocado_2006a}, Fe-Rh~\cite{annaorazov_1992a,annaorazov_1996a}).  Several clues as to how to compare materials emerge.  Despite having a sharp, first order transition, MnAs scores poorly as its {\Tadmax} is very low for $\mu_{0}${\Hmax}$<$ 2.5~Tesla, due to the large transition hysteresis.  The second order ferromagnets, Gd and La(Fe,Co)Si appear to the left of the plot, their entropy change values being limited by the lack of sharpness of their phase transition.  First order materials (hydrogenated La-Fe-Si and Mn-Fe-P-based) are slightly to the right. Fe-Rh is a notable exception in that its {\Tadmax} is remarkably high as highlighted by Zverev et al.~\cite{zverev_2010a}. 

Our focus will be on the proximity of $\partial T_{c} / \partial H$ to $\eta$.  The high value of {\Tadmax} in Fe-Rh can be understood in terms of its value of $\mu_{0}^{-1}\partial T_{c} / \partial H \sim$-7~KT$^{-1}$.  Let us use realistic values in Eq.~\ref{eq_dTdHmax} of $M_{\rm sat}\sim120$~Am$^{2}$kg$^{-1}$, a formula weight of 159~a.m.u. and $C_{p}\sim$~6R JK$^{-1}$mol$^{-1}$ due to phonon modes for the two atoms in this binary system, away from the phase transition.  For $\mu_{0}${\Hmax}=2~Tesla, we have a theoretical optimum value of {\Tadmax} of 15~K when $\mu_{0}^{-1}\partial T_{c} / \partial H=\mu_{0}^{-1}\eta_{\rm FeRh}=-7.7$~KT$^{-1}$.  

What therefore sets Fe-Rh apart from other alloys is that its $\partial T_{c} / \partial H$ value is so close to $\eta$.   By contrast if we examine the La-Fe-Si compounds, a similar analysis yields an optimum {\Tadmax} of 13~K in 2~Tesla if $\mu_{0}^{-1}\partial T_{c} / \partial H=6.5$~KT$^{-1}$.  This, and a similar calculation for Mn-Fe-P are shown in Figure~\ref{DTDSFig} by the filled circles.  Actual {\Tadmax} values of around 4-7~K in 2~Tesla are therefore due to $\mu_{0}^{-1}\partial T_{c} / \partial H$ values of around 4-5~KT$^{-1}$, which are typical of Curie ferromagnets.  The largest {\Tadmax} that could be expected given such experimental values of $\partial T_{c} / \partial H$ are shown by open circles.  Fe-Rh, Mn-Fe-P and La-Fe-Si lie along the same line for $\vert \partial T_{c} / \partial H \vert<\eta$ as we have used $M_{\rm sat}\sim120$~Am$^{2}$kg$^{-1}$ for each.  We note that, of the three, Fe-Rh is by far the closest to its $\eta$ point.  So an open question is how to use (nano-)structure to {\it increase} the $\partial T_{c} / \partial H$ values of existing materials so as to raise {\Tadmax} and the {\Tadmax}{\DSmax} product.

At present, first order La-Fe-Si- or Mn-Fe-P-based materials offer the best combination of magnetothermal performance and raw material cost.  They have field-induced critical points, where low hysteresis, first order ferromagnetism gives way to a continuous Curie transition~\cite{fujita_2003a,trung_2009a}.  This regime is ideal for applications.  Investigation of metamagnetic critical points, both in Curie ferromagnets such as MnFe(P,Si)\cite{dung_2011b}, CoMnGe~\cite{hamer_2009a,trung_2010b} and La-Fe-Si~\cite{morrison_2010b} and in tricritical antiferromagnets~\cite{barcza_2010a} is yielding fundamental insight.   A key question is how to use critical fluctuations to mediate low hysteresis first order phase transition behaviour, as discussed in early work on Fe$_{2}$P~\cite{lundgren_1978a}.  

Along with tuning refrigerants close to a critical point, what other options are available?  Refrigerant material systems such as MnFe(P,Si), Gd$_{5}$(Si,Ge)$_{4}$ and La(Fe,Si)$_{13}$ exhibit substantially sub-maximal values of the {\Tadmax}{\DSmax} product; in 2~Tesla they cluster around {\Tad}{\DS}=100~Jkg$^{-1}$.  With $M_{\rm sat}\sim$120~Am$^{2}$kg$^{-1}$, we might expect up to 240~Jkg$^{-1}$ from the sum rule, even if $\partial T_{c} / \partial H > \eta$.  Similar comparisons of {\DSmax} with the magnetic entropy limit have previously been made, for example in the MnFe(P,As)~\cite{tegus_2002a}.  However, the finite magnetisation in the paramagnetic state reduces the observed field-induced entropy change.  (The above model assumes a digital switch of magnetisation.) In manganites, short range magnetic ordering above the Curie temperature has in fact been quantified~\cite{jia_2006a}.  In Mn-Fe-P-based alloys it has been explained by the presence of fluctuating moments near the Curie temperature~\cite{wilkinson_1989a}.  The supression of critical fluctuations might help to raise the entropy change, but it may also make the observation of low hysteresis difficult~\cite{lundgren_1978a}.

The above comparison therefore hints at room for improving first order MCEs, if $\partial T_{c} / \partial H$ can be tuned, and/or if the initial magnetisation away from $T_{c}$ can be suppressed.  It also highlights another reason for the high performance of Fe-Rh and of other antiferromagnets (AFMs) such as MnGa$_{3}$C~\cite{tohei_2003a}.  Namely, metamagnetic antiferromagnets with sufficient anisotropy can exhibit a field-induced jump in magnetisation that is almost equal to $M_{\rm sat}$, since only a small magnetisation tail is often present in the AFM state.  They can often exhibit highly tunable metamagnetism based on giant magneto-elastic interactions~\cite{barcza_2010a} and offer a testing ground for theory, which can now predict metamagnetism successfully based upon structural information~\cite{gercsi_2011a}.   An ability to reduce $\partial T_{c} / \partial H$ may be necessary since this parameter can be much greater than $\eta$~\cite{sandeman_2006a}.  Lastly, changes in a particular site moment across a sharp magnetic transition are also of interest.  They already lead to enhanced MCE, both in Fe-Rh (where the Rh moment is almost 1~$\mu_B$ in the ferromagnetic state~\cite{shirane_1963a,shirane_1964a}) and in Mn-Fe-P-based materials (where the Mn and Fe sites are crystallographically distinct~\cite{dung_2011a}).  

Magnetic anisotropy is the final atomic- or nano-scale parameter at the material physicist's disposal.  It has been shown that anisotropic materials yield increased {\DSmax}~\cite{bennett_1993a} and that the choice of axis along/about which to apply/rotate a magnetic field can greatly influence the MCE in spin reorientation compounds~\cite{lima_2004a}.   Indeed, spin reorientation is a less-explored route to obtaining MCEs, typically examined in rare earth intermetallics~\cite{kuzmin_1993a}.  Hard ferrites have also come under renewed investigation~\cite{lobue_2012a}, but the largest effects are still seen in rare earth-based materials~\cite{nikitin_2010a},  This is not surprising given that they have the largest thermally-varying anisotropy constants, as is required in an analogous way to the large $\partial M / \partial T$ required for paraprocess-based MCE (Eq.~\ref{DSEq} and Ref.~\cite{kuzmin_2007b}).    However, recent developments are shedding light on how to: tailor nano-scale materials to tune magnetocaloric effects through anisotropy control in ferromagnetic nanostructures~\cite{franco_2008a}; adjust the shape and height of the $\Delta S(T)$ response in Gd/W heterostructures~\cite{miller_2010a} and manganite superlattices~\cite{zhang_2011a}; and drive new kinds of metamagnetism in metamagnetic heterostructures~\cite{sahoo_2003a}.  Predictive control of the magnetothermal response can be difficult~\cite{mukherjee_2009a} but will provide interaction between the nano-magnetism, thin film and magnetocaloric communities. 


\section{New systems to aid deployment}
\label{sec_deployment}
The previous sections considered how atomic- or nano-scale properties may adjust {\Tadmax} and {\DSmax} towards the theoretical limits more closely exhibited by Fe-Rh.   The grouping of Mn-Fe-P- and La-Fe-Si-based materials in a similar region of Figure~\ref{DTDSFig} highlights the role of materials physics in fine-tuning the {\DS} and {\Tad} responses in these materials and the need to feedback the output of non-magnetothermal characterisation to physical modelling.  Synthesis methods and resulting microstructure have a large role to play in establishing new material systems as they greatly affect the presence of secondary phases~\cite{lyubina_2008a,morrison_2010a}, homogeneity~\cite{barcza_2011a} and internal strain~\cite{morrison_2008a}.  These in turn can affect machinability, shapeability, thermal conductivity and other important interdisciplinary variables. 

A good example is La-Fe-Si.  It is the most trialled non-Gd-based refrigerant.   To fabricate flat plates for regenerators, several hurdles have to be overcome.  The plates need to have a mm or sub-mm thickness, and so structural integrity is a relevant property.  Machinability can be impaired by the presence of magnetovolume effects at the room temperature Curie transition, and so new techniques have been developed, typically using insight gained from studies of the metallurgical phase diagram.  A thermal decomposition and recombination (TDR) process has enabled machining in a predominantly Fe-containing state, followed by heat treatment to achieve the desired composition~\cite{katter_2010a}.  A second new process allows solid hydrogenation of desired shapes without decrepitation, by controlling the rate of change of temperature during hydrogenation~\cite{katter_2011a}.  Lastly, microscopy studies are enabling greater control of the two-phase nature of this system, which forms by a peritectic reaction~\cite{liu_2011a}.  A future goal will  be to combine growing knowledge of the phase diagram with advanced processing techniques in order to yield desired shapes reproducibly.   A similar statement may be made about (Mn,Fe)$_{2}$(P,Ge) and (Mn,Fe)$_{2}$(P,Si) which have been proposed for use in regenerators and for which control over morphology and composition will be key, since in the latter, $T_{c}$ varies strongly with Si content~\cite{dung_2011a}.

Like La-Fe-Si and Mn-Fe-P, the manganites can sit close to a critical point.  Indeed, the  first or second order nature of the ferromagnetic phase transition is debated in particular compositions~\cite{loudon_2006a}.  Although the high heat capacity (per kg) of the manganites somewhat limits the observed {\Tad}~\cite{pecharsky_2001a} --- indeed it sits more towards the bottom left of Figure~\ref{DTDSFig} --- the cost and shapeability of this material system offers prototyping potential.  Regenerator blocks of stacked plates have been developed that demonstrate the shapeability of tape cast material down to 0.3~mm~\cite{engelbrecht_2011b}.  The thermal conductivity of this system is the lowest of the three materials in this section and so efforts to investigate its impact and methods of tuning it will be of relevance to future deployment.  Simultaneous control of the magneto-elastic interaction, oxygenation and disorder will be required in order to harness the relatively high {\Tad} sometimes observed in 1~Tesla~\cite{dinesen_2005a}.  The position of manganites relative to other first order materials in a 1~Tesla version of Fig.~\ref{DTDSFig} may then improve.

\section{Conclusions}
From idealised first order magnetothermal properties, this Viewpoint has attempted to identify key challenges in the search for new magnetic refrigerants, materials to be established and tuned by varying structure at several lengthscales.  Limits on {\Tadmax} and {\DSmax} have been found and experimental data have been used to compare known refrigerants on an Ashby-type map.  First order materials are found to already provide better properties at a single temperature point than Gd.  We note that operation close to a critical point is preferable in order to reduce hysteresis to a minimum.

Routes to further optimise core magnetothermal properties at the atomic- and nano-scale have been suggested, focussing on systems close to critical points, those with step changes in site moments, and those in which $\mu_{0}^{-1}\partial T_{c} / \partial H$ can be tuned towards a value of about 7~KT$^{-1}$.  Lastly, examples of microstructuring in refrigerants currently being trialled give us valuable clues as to how to optimise both established compositions and future materials for inexpensive near-term deployment in magnetic cooling engines.  



%
%

%

\section{Acknowledgments}
The research leading to these results has received funding from the
European Community\textquoteright{}s 7th Framework Programme under
grant agreement 214864 ``SSEEC" and was supported by EPSRC grant EP/G060940/1. K.G.S. acknowledges financial support from
The Royal Society and useful discussions with V. Basso, L.F. Cohen, N. Demspey, Z. Gercsi, O. Gutfleisch, M. Katter, M. Lo Bue, A. Pastore, J.B. Staunton and N. Wilson.

\bibliographystyle{model1a-num-names}
\bibliography{RefrigerantViewpoint.bib}


\end{document}